# Generation and modulation of multiple 2D bulk photovoltaic effects in space-time reversal asymmetric 2H-FeCl$_2$

Liang Liu[1,2], Xiaolin Li[1], Luping Du[3], Xi Zhang[1,4,†]

[1] Institute of Nanosurface Science and Engineering, Guangdong Provincial Key Laboratory of Micro/Nano Optomechatronics Engineering, Shenzhen University, Shenzhen 518060, China

[2] School of Physics, State Key Laboratory for Crystal Materials, Shandong University, Jinan 250100, China

[3] Nanophotonics Research Centre, Institute of Microscale Optoelectronics, Shenzhen University, Shenzhen 518060, China

[4] Research Center of Plasma Medical Technology, Shenzhen University, Shenzhen 518060, China

Corresponding author. E-mail: [†]zh0005xi@szu.edu.cn

**Abstract**

The two-dimensional (2D) bulk photovoltaic effect (BPVE) is a cornerstone for future highly efficient 2D solar cells and optoelectronics. The ferromagnetic semiconductor 2H-FeCl$_2$ is shown to realize a new type of BPVE in which spatial inversion (P), time reversal (T), and space-time reversal (PT) symmetries are broken (PT-broken). Using density functional theory and perturbation theory, we show that 2H-FeCl$_2$ exhibits giant photocurrents, photo-spin-currents, and photo-orbital-currents under illumination by linearly polarized light. The injection-like and shift-like photocurrents coexist and propagate in different directions. The material also demonstrates substantial photoconductance, photo-spin-conductance, and photo-orbital-conductance, with magnitudes up to 4650 (nm·µA/V$^2$), 4620 (nm·µA/V$^2$ $\hbar$/2e), and 6450 (nm·µA/V$^2$ $\hbar$/e), respectively. Furthermore, the injection-currents, shift-spin-currents, and shift-orbital-currents can be readily switched via rotating the magnetizations of 2H-FeCl$_2$. These results demonstrate the superior performance and intriguing control of a new type of BPVE in 2H-FeCl$_2$.

## 1. Introduction

Bulk photovoltaic effects (BPVE) convert light into steady currents in single homogeneous material, displaying great potential in solar energy, information technologies, optical spectroscopes, etc. [1-10]. These effects have been observed in several bulk ferroelectric perovskites such as LiNbO$_3$ and BiFeO$_3$ [3-6, 11-15], in which the spatial inversion (P) symmetry is broken. Recently, BPVE were also observed in P asymmetric two-dimensional (2D) systems such as 2H-WSe$_2$, MoSe$_2$, CuInP$_2$S$_6$, and MoS$_2$ nanotube [7-10, 16-17], in which the shift-currents dominated the photocurrents under the illuminations of linearly polarized incident light. Since the time reversal (T) symmetries are preserved, these 2D systems belong to the class 1 BPVE (see Table 1).

Breaking the T symmetry of semiconductors is essential to manipulate the spin, orbital and valley polarizations, which are the keys in spintronics and valleytronics. [18-22] T asymmetry thus should significantly enrich the applications of 2D BPVE. Unfortunately, T asymmetry is much harder to achieve than P asymmetry. One tangible route is to apply a large magnetic field to induce Zeeman effect and break T symmetry. [21-26] This might be inefficient since the Zeeman effects usually on the magnitudes of several meVs per Tesla, [23-26] which is too weak for realistic applications. Magnetic semiconductors intrinsically break T symmetry with exchange field, providing significant observable effects. [18, 27-33] Owing to the continuous efforts, researchers found that the 2D anti-ferromagnet (AFM) $CrI_3$ [7], $MnBi_2Te_4$ [9, 34] and $MnPSe_3$ [35] are supposed to break both P and T symmetries and achieve 2D magnetic BPVEs. Since the AFM systems remain unchanged under the combination operations of inversion and time reversal, they preserve the PT symmetry and the BPVEs are class 2 (Table 1). Besides, the P, T asymmetries in these two systems are only enabled by the weak AFM orders, which are not very robust against thermal fluctuations. Consequently, photocurrents in these AFM are relatively weak, and their working temperature (Tc) are rather low (~45 K for $CrI_3$, ~25 K for $MnBi_2Te_4$ and ~80 K for $MnPSe_3$). Therefore, realizing sizable and stable T asymmetries in 2D BPVEs is highly desired for the fundamental studies and realistic applications.

In this work, we propose that the 2D ferromagnet (FM) 2H-$FeCl_2$ is an ideal system to realize room temperature 2D BPVEs. Based on density functional theory (DFT), the spontaneous breaking of P, T, and PT symmetries in 2H-$FeCl_2$ are justified, suggesting 2H-$FeCl_2$ as the candidate for the class 3 BPVE. The symmetry structure of 2H-$FeCl_2$ enables synchronized injection-type and shift-type photocurrents via illuminations of linearly polarized light. The effects of the two distinct bulk photocurrents propagates in different directions and the nonlinear photoconductance exceeds 4650 (nm·μA/V$^2$), which is 1~2 orders larger than known BPVEs. Furthermore, due to large exchange fields and nontrivial orbital texture in the band structure, photocurrents in 2H-$FeCl_2$ are also spin-polarized and orbital-polarized. Through rotating magnetization orientations in 2H-$FeCl_2$, one may switch the injection-currents, shift-spin-currents, and shift-orbital-currents. These findings indicate that 2H-$FeCl_2$ is suitable to realize switchable, sizable, and room temperature 2D BPVEs which may merit future applications and studies on 2D energy devices, spintronics and orbitronics.

**Table 1** The symmetry classification, mechanism, and typical platforms for the nonlinear photocurrents.

| Class | Symmetry | BPVE mechanism | concrete 2D crystal |
| --- | --- | --- | --- |
| 1 | P-broken<br><br>T-conserved | Linear light: shift<br><br>Circular light: injection | $MoS_2$, [16-17, 36]<br>$MoSe_2$, $CuInP_2S_6$ [8] |

| 2 | P-broken  T-broken  PT-conserved | Linear light: injection  Circular light: shift + injection | Bilayer CrI$_3$ [7], bilayer MnBi$_2$Te$_4$ [9, 34], 1T-FeCl$_2$ [37]  Monolayer MnPSe$_3$ [35] |
|---|---|---|---|
| 3 | P-broken  T-broken  PT-broken | Linear light: shift + injection  Circular light: shift + injection | Monolayer 2H-FeCl$_2$ (this work) |

## 2. Computational details

The relativistic DFT calculations on 2H-FeCl$_2$ were done within the Vienna atomic simulation pack (VASP). [38] The Hilbert space for DFT calculations were built on the basis of projected augmentation plane waves (PAW) [39] truncated at 300 eV, and The Perdew−Burke−Ernzerh (PBE) form functionals on generalized gradient approximation (GGA) [40] with considerations on relativistic effects and spin-orbital-coupling were used to describe the exchange-correlation interactions between electrons. Gamma-centered 50×50×1 mesh sampled via Monkhorst-Pack scheme [41] is used to conduct the integrations on Brillouin zone. The numerical calculations for photoconductances are obtained from denser k-mesh 1600×1600×1 and the results differ from the ones with 800×800×1 by less than 5%. The Hamiltonian is built from the first-principle wavefunctions and then interpolated with WANNIER90 package. [42-44] The calculations on bulk photoconductance were based on the Wannier Hamiltonian, up to the relaxation time approximation in which the relaxation time $\tau$ is set to 0.2 ps.

## 3. Geometric and electronic structures

Figure 1(a) shows the top view of 2H-FeCl$_2$ monolayer, where Fe$^{2+}$ and Cl$^-$ ions form honeycomb structure. The mirror symmetry is obvious with the mirror plane normal to the *x*-axis, indicating that the *x*-direction is a nonpolar direction of 2H-FeCl$_2$ lattice. Fig. 1b shows the local atomic structure, displaying the point symmetry D$_{3h}$ which lacks P symmetry. The D$_{3h}$ type inversion asymmetries have been experimentally observed in several 2D systems, e.g., 2H-MoS$_2$, 2H-MoSe$_2$, 2H-WSe$_2$. [45-50] The kinetical and thermal stabilities of 2H-FeCl$_2$ were reported in our former work, [51] it is thus promising to realize 2H-FeCl$_2$ in laboratory by the means of mechanical exfoliations, chemical vapor deposition, etc. Distinct from conventional 2H phase 2D systems (MoS$_2$, WSe$_2$, etc.), 2H-FeCl$_2$ hosts open d-shells in Fe$^{2+}$ ions, which give rise to the intrinsic 2D magnetizations. According to our former works, [51] the magnetic moments of Fe$^{2+}$ ions is 4 μB. The easy axis is along the out-of-plane direction with magnetocrystalline anisotropy energy of ~50 μeV per Fe$^{2+}$ ion. And the strong ferromagnetic exchange couplings between Fe$^{2+}$ lead to the stable long-range ferromagnetic orders up to the Curie temperature of ~930 K. Since the time reversal operations which flip all the magnetic moments will cause inequality, as sketched in Fig. 1(b), the breaking of T symmetry in 2H-FeCl$_2$ should be stable up to room temperature.

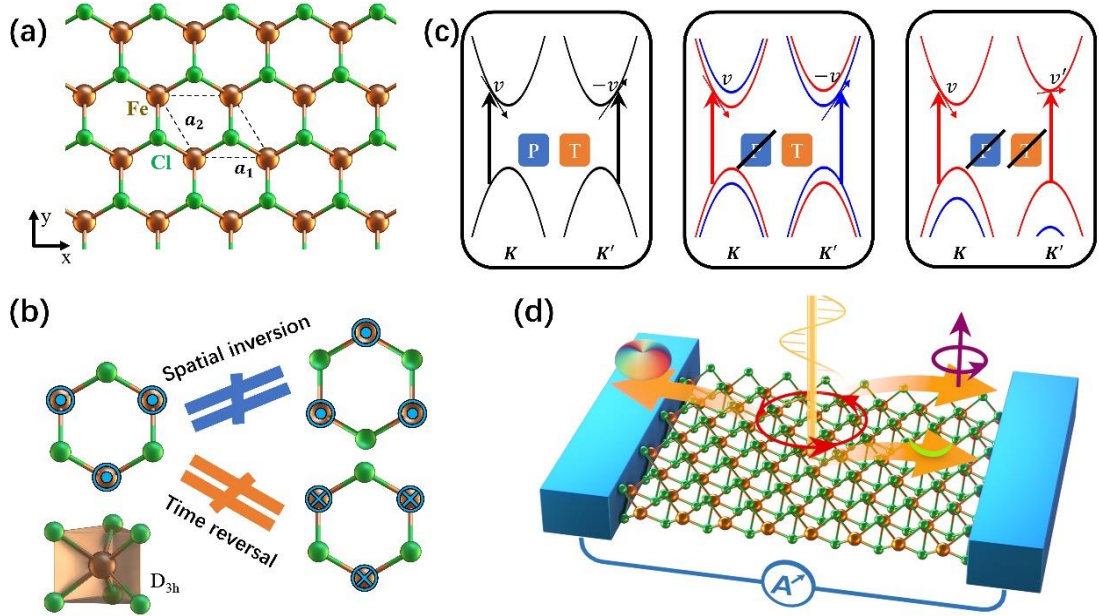

**Fig. 1** Structure of class 3 BPVEs in 2H-FeCl$_2$. **(a)** Geometry structure of 2H-FeCl$_2$. Green and brown balls denote Cl and Fe atoms. Dashed lines outline the periodic cell. **(b)** Local structures of 2H-FeCl$_2$. Blue dots denote the magnetic moments of Fe atoms along $+z$-direction. Blue crosses denote the magnetic moments along $-z$-direction. Left part is the original structure, right part involves transformed structures through spatial inversion or time reversal. **(c)** The typical valley ($K$ and $K$' = $-K$) electronic structures in systems with both P and T symmetries, with only T symmetry, and without P or T symmetries. Black, red, and blue curves denote the spin compensated, spin up and spin down states. Arrows denote the electron-hole generation processes via absorbing photons. **(d)** The prototype of 2D BPVE device, in which steady currents, spin-currents and orbital-currents are generated by applying light.

Figure 1(c) shows the symmetry-enforced topologies in electronic structures. $K$ and $K$' are the valleys with opposite crystal momentums, i.e., $K$' = $-K$. In the case that P and T symmetries are both preserved, electronic bands must be degenerated everywhere, and the group velocities in $K$ and $K$' are exactly opposite, and the photocurrents vanish ideally. The second part of Fig. 1(c) shows the case where P is broken but T is preserved, corresponding to the class 1 BPVEs. On opposite momentums, the energies are degenerate, but the spin tunnels, orbital momentums and Berry curvatures may acquire opposite values. Hence, the pure spin-currents, orbital-currents, and shift charge-currents may be induced under photon pumping, as suggested by recent studies. [9, 35, 52] The last part of Fig. 1(c) shows the case where both P and T are broken, which is the case of class 3 BPVE in 2H-FeCl$_2$. Apparently, the electrons pumped by photons with the same energy acquire different group velocities in opposite valleys, so that the injection currents emerge. Furthermore, the spin currents and shift currents are still possible in the last case due to the P asymmetry. Therefore, 2H-FeCl$_2$ with P and T violated coherently should present abundant optoelectrical responses.

Figure 1(d) sketches the devices for the observations on the optoelectrical responses in 2H-FeCl$_2$. Under illuminations with proper frequencies, 2H-FeCl$_2$ generates steady currents. Since the generations of currents depends on the band structures in which the spin and orbital polarizations are presented, it is possible to generate and modulate spin and orbital currents in 2H-FeCl$_2$, displaying benefits for 2D spintronics and orbitronics.

Figure 2(a) and (b) show the band structures of 2H-FeCl$_2$ with and without the effects from spin-orbital-couplings (SOC). The magnetic moments of 2H-FeCl$_2$ in this case are in +$z$ direction, which is the energetically favored state. Figure 2(d) shows the irreducible Brillouin zone (BZ) and several high symmetric k-points in BZ. For both cases in Fig. 2, the bands in spin up tunnels have energies much lower than spin down tunnels, the exchange fields are thus significantly large. The Fermi level is in the gap between spin up states, so that the six d-electrons of each Fe$^{2+}$ have occupied the whole spin up d-shell and one spin down states, and 2H-FeCl$_2$ exhibits a novel ferromagnetic semiconductor with magnetic moment of 4 μB per unit cell, consistent with previous studies. [51]

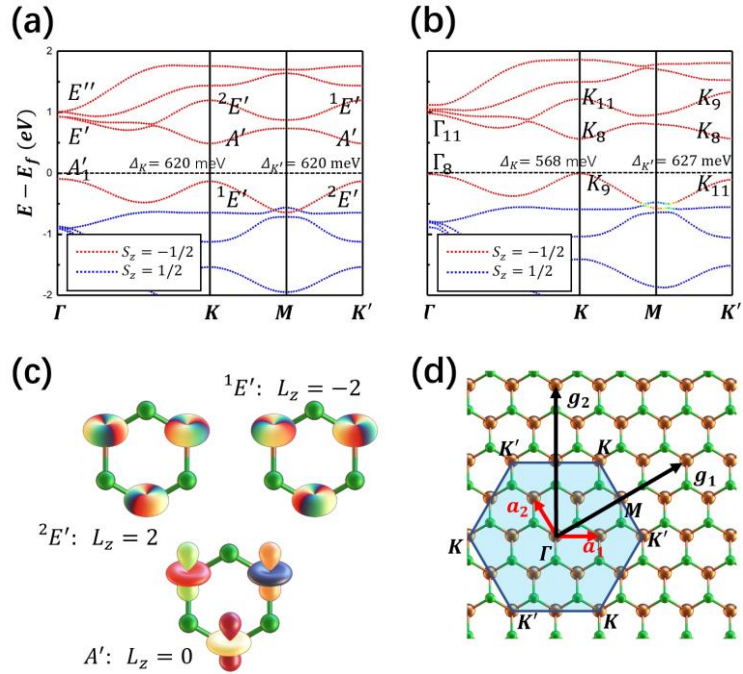

**Fig. 2** Electronic structures of 2H-FeCl$_2$. **(a)** Band structure without SOC and **(b)** with SOC. Red and blue lines represent spin down and up states. The irreducible representations of several states in high symmetric k-points were marked. Energy gaps at $K$ and $K'$ were defined as $\Delta_K$ and $\Delta_{K'}$. **(c)** The atomic wave functions of irreducible representations of states in FeCl$_2$. Color denotes the phases of wavefunctions. **(d)** Irreducible Brillouin zone of 2H-FeCl$_2$. $a_1$, $a_2$ are lattice vectors. $g_1$, $g_2$ are reciprocal lattice vectors.

The states near energy gaps are valley states. It is apparent from the band structure in Fig. 2 that two distinct types of valleys $K$ and $K'$ are presented in 2H-FeCl$_2$. The essential energy gaps in 2H-FeCl$_2$ come from the D$_{3h}$ triangular-prism crystal fields shown in Fig. 1(b). For the case without SOC [Fig. 2(a)], the highest valence states at $K$ and $K'$ presents $^1E'$ and $^2E'$ irreducible representations, corresponding to the $d$-orbitals of Fe$^{2+}$ with $L_z = -2$ and $+2$ [sketched in Fig. 2(c)]. The lowest conducting states at $K$ and $K'$ all have the identity representation $A'$, corresponding to the $d_{z^2}$ orbitals of Fe$^{2+}$ [also sketched in Fig. 2(c)]. Without SOC, there is latent symmetry operation $T \times S$ in system, where $S$ reverses the spin. $T \times S$ flips the signs of momentums but preserve the spin orientations. Therefore, the energy levels in $K$ and $K'$ are aligned, with energy gaps $\Delta_K = \Delta_{K'}$ = 620 meV (GGA level).

The incorporation of SOC violates the $T \times S$ symmetry, and energy levels in $K$ and $K'$ no longer align with each other, leading to the energy bands topologies like the last part of Fig. 1(c). Specifically, the lowest conducting states of FeCl$_2$ still have the same energies, while the highest valence states shift up and down for $K$ and $K'$ [Fig. 2(b)]. Then the energy gaps in $K$ and $K'$ become $\Delta_K$ = 568 meV and $\Delta_{K'}$ = 627 meV, and the valley splitting energy is ~60 meV, comparable to typical valleytronic systems. [53-54] It is straight to understand these energy shifts in terms of perturbation formalism. Effects of SOC can be captured with following terms:

$$H \propto -\xi \left( L^z S^z + \frac{(L^+ S^- + L^- S^+)}{2} \right) , \quad (1)$$

where $L^{+(-)}, S^{+(-)}$ are orbital and spin raising (lowing) operators. $\xi > 0$ is the strength of SOC. Up to the first order, the energy shifts due to SOC is $\Delta E = -\xi \langle \psi | L^z S^z | \psi \rangle$. In the lowest conducting states with $L^z=0$, the corresponding SOC energies are zero. On the other hand, Fig. 2 shows that the highest valence states have $S_z = -\hbar/2$ and $L_z = \pm 2\hbar$ for $K$ and $K'$, so that these states have energies ascended and descended due to the SOC.

More importantly, since the spin polarization directions in ferromagnet are amenable to external magnetic field, it is accessible to reverse the $S_z$ of valence states, leading to the switching of valley polarization in 2H-FeCl$_2$. To be specific, when the magnetic moments oriented to $+z$, the red dotted states in Fig. 2(b) have $S_z \approx \hbar/2$, and the system is valley polarized with $\Delta_{K'} > \Delta_K$. once the magnetic moments rotated to $x/y$ (parallel to atomic plane), we have $S_z \approx 0$, and the valley polarizations are vanishingly small; for magnetic moments pointing to $-z$, the system exhibits valley polarizations with $\Delta_{K'} < \Delta_K$. Hence, the valley polarizations can be continuously tuned via rotating the magnetic moments. The efficient manipulations on polarizations in the band structure is essential to realize extraordinary controllable optoelectronic responses.

## 4. 2D bulk photocurrents, photo-spin-currents and photo-orbital currents in 2H-FeCl$_2$

To directly observe the 2D BPVE, the nonlinear photoconductance is calculated on the basis of the 2$^{nd}$ order nonlinear optoelectrical responses. The 2$^{nd}$ order photocurrents in 2H-FeCl$_2$ produced by incident light with frequency $\omega$ can be formally expressed as

$$j^{\gamma}_{C/S/L} = \text{Re} \sum_{\alpha\beta} \sigma^{\gamma:\alpha\beta}_{C/S/L} E^{\alpha}(\omega) E^{\beta*}(\omega) \ . \qquad (2)$$

Here the superscripts $\alpha$, $\beta$ and $\gamma$ can be $x$ or $y$ or $z$, labeling the spatial directions of corresponding vectors and tensors, while the subscripts $C$ or $S$ or $L$ label the type of currents. $j^{\gamma}_C$, $j^{\gamma}_S$, and $j^{\gamma}_L$ are the photocurrents, photo-spin-currents, and photo-orbital-currents propagating along $\gamma$-direction. $E^{\alpha}$ denotes the electric field component of polarized light in $\alpha$-direction. $\sigma^{\gamma:\alpha\beta}_C$, $\sigma^{\gamma:\alpha\beta}_S$, and $\sigma^{\gamma:\alpha\beta}_L$ are the photoconductance, photo-spin-conductance, and photo-orbital-conductance tensors. By virtue of the three-fold rotation symmetry in 2H-FeCl$_2$, the following relations hold for the three kinds of photoconductance components: $\sigma^{x:xx}_{C/S/L} = -\sigma^{x:yy}_{C/S/L} = -\sigma^{y:xy}_{C/S/L} = -\sigma^{y:yx}_{C/S/L}$, $\sigma^{y:yy}_{C/S/L} = -\sigma^{y:xx}_{C/S/L} = -\sigma^{x:xy}_{C/S/L} = -\sigma^{x:yx}_{C/S/L}$, that is, only two components $\sigma^{x:xx}_{C/S/L}$ and $\sigma^{y:yy}_{C/S/L}$ are independent. Considering the linearly polarized beams with electric field confined to $x$-direction, the first independent component $\sigma^{x:xx}_{C/S/L}$ describes the photocurrents transporting along $x$-direction which is also the nonpolar direction of 2H-FeCl$_2$, while $\sigma^{y:xx}_{C/S/L}$ describes the photocurrents along $y$-direction.

The conductance can be further calculated via Kubo's formula [7, 55-56]:

$$\sigma^{\gamma:\alpha\beta}_{C/S/L} = \sum_{\Omega=\pm\hbar\omega} \frac{2|e|^3}{S\Omega^2} \sum_{lmn,k} f_{ln} \frac{\langle n,k|v^{\alpha}_C|l,k\rangle \langle l,k|v^{\beta}_C|m,k\rangle \langle m,k|v^{\gamma}_{C/S/L}|n,k\rangle}{(E_{nk}-E_{mk}-i\eta)(E_{nk}-E_{lk}-\Omega)} \qquad (3)$$

where the newly introduced subscripts of $l$, $m$, $n$ represent the index of bands, $k$ represents the index of $k$-points on the irreducible Brillouin zone. $\Omega$ denotes the photon energy, $S$ is the area of unit cell, $|n,k\rangle$ is the $n^{th}$ eigenstate with energy $E_{nk}$. $v^{\alpha}_C$ is the $\alpha$-component of velocity operator. $v^{\alpha}_S = (S_z v^{\alpha}_C + v^{\alpha}_C S_z)/2$ and $v^{\alpha}_L = (L_z v^{\alpha}_C + v^{\alpha}_C L_z)/2$ are the spin-resolved and orbital-resolved velocity operators. [52, 57-58] $f_l$ is the occupation number satisfying the Fermi-Dirac distribution: $f_l = (1+\exp\beta(E_l - E_f))^{-1}$ in which $E_f$ is the Fermi energy and $\beta=1/kT$ is inversed temperature. $f_{ln} = f_l - f_n$ is the occupation difference. $\eta=\hbar/\tau$ corresponds to the relaxation energy, and $\tau$ is the relaxation time. Since the calculated photocurrents may be proportional to the relaxation time, [7] a relatively low relaxation time $\tau = 0.2$ ps is utilized to avoid overestimations. Usually, the relaxation time of clear 2D crystals can reach ~1 ps and even ~1 ns, [59] the 0.2 ps used here is thus rather conservative. The thickness of 2D monolayer is ambiguously defined, the effective thickness $L = 1$ nm is used in the following calculations. Because the averaged photoconductance in Eq. (3) is inversely proportional to the thickness, the setting of $L = 1$nm (larger than the thickness of conventional vdW monolayer such as grapheme and MoS$_2$) should not overestimate the magnitudes.

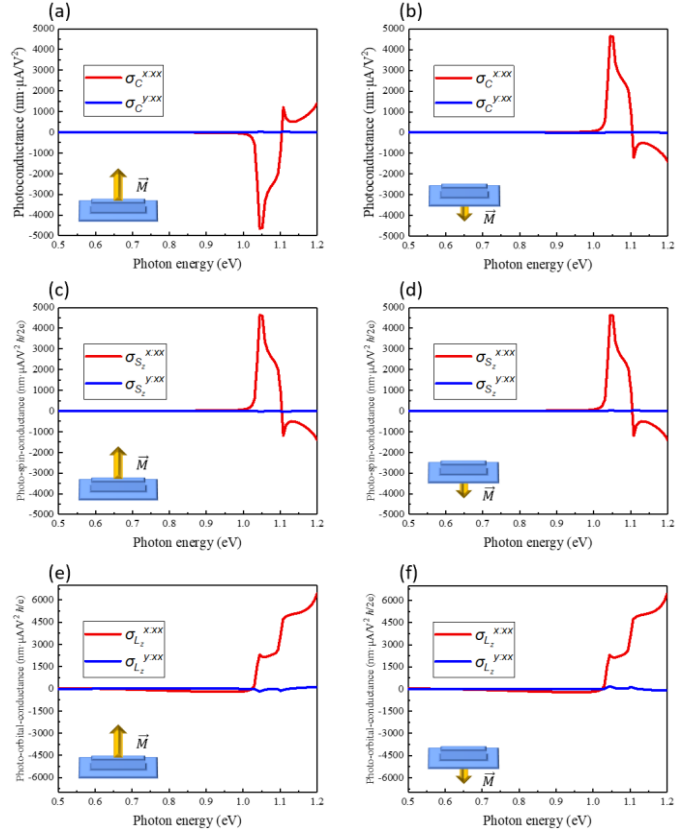

**Fig. 3** Calculated 2$^{nd}$ order photoconductance **(a, b)**, photo-spin-conductance **(b, c)**, and photo-orbital-conductance **(e, f)**. The magnetizations of 2H-FeCl$_2$ in (a), (c), and (e) are in +$z$-direction. The magnetizations in (b), (d), and (f) are in −$z$-direction.

Figures 3(a) and (b) show the photoconductance components $\sigma_C^{x;xx}$ and $\sigma_C^{y;xx}$ in 2H-FeCl$_2$ with magnetizations along ±$z$-directions. For +$z$ magnetization [Fig. 3(a)], $\sigma_C^{x;xx}$ acquires the maximal magnitude −4650 (nm·μA/V$^2$) at photon energy 1.05 eV, then $\sigma_C^{x;xx}$ changes sign and keeps increasing with photon energies. On the other hand, the maximal value of $\sigma_C^{y;xx}$ is 25 (nm·μA/V$^2$) at photon energy 1.11 eV, which is two magnitude orders smaller than $\sigma_C^{x;xx}$. Therefore, under the illuminations of linearly polarized light, the photocurrents mainly running in the $x$-direction which is also the nonpolar direction of 2H-FeCl$_2$ lattice. Comparing to the class 1 BPVEs that are hosted by the nonmagnetic semiconductors, the PT-broken symmetry in 2H-FeCl$_2$ enables both the shift-like and injection-like photocurrents, revealing potentials for much larger BPVE efficiency. The photoconductance of ~4650 nm·μA/V$^2$ shown here is equivalent to a photo-responsivity of ~1752 mA/W, which shows better performance than the 2D ferroelectric CuInP$_2$S$_6$ (~3 mA/W) [8], the MoS2 nanotube (~ 100 mA/W) [16], and the strained MoS2 (tens of mA/W) [17, 36]. Comparing the class 2 BPVEs of PT-antiferromagnet, the 2H-FeCl$_2$ also show better photon-to-current conversions than the bilayer CrI$_3$ (~200 nm·μA/V$^2$) and bilayer MnBi$_2$Te$_4$ (~30 nm·μA/V$^2$). So, the PT-broken symmetry is suitable for improving the efficiency of 2D BPVE.

Figure 3(d) shows that the $\sigma_C^{x;xx}$ in system with magnetization along $-z$-direction is exactly the reversion of the one in system with magnetizations along $+z$ direction [Fig. 3(b)], and the strongest $\sigma_C^{x;xx}$ is 4650 (nm·mA/V$^2$) at photon energy 1.05 eV. Noticing the facts that the two cases with magnetizations in $\pm z$-directions are the time reversal partners, it can be speculated that the $\sigma_C^{x;xx}$ is dominated by injection-like contributions since $\sigma_C^{x;xx}$ can be reversed by the time reversal operations. In addition, $\sigma_C^{y;xx}$ of 2H-FeCl$_2$ with $-z$-magnetization is the same as the case with $+z$-magnetization, showing maximal value 25 (nm·μA/V$^2$) at photon energy 1.11 eV. Therefore, the $\sigma_C^{y;xx}$ is independent on the time reversal operations, showing features of shift-like contributions which come from the nontrivial Berry curvatures.

Figure 3(c) and (d) show the photo-spin-conductance. The evolutions are close to the photoconductance displayed in Fig. 3(b). The largest $\sigma_{Sz}^{x;xx}$ is 4620 (nm·μA/V$^2$ $\hbar/2e$) at photon energy 1.05 eV, representing the injection-like photo-spin-currents. $\sigma_{Sz}^{x;xx}$ cannot be reversed by flipping the magnetizations. The reason is straight: the time reversal operation changes the sign of spins and photocurrents simultaneously, then the whole signs of photo-spin-currents preserve. The similarities between photo-spin-conductance and photoconductance reveals that the photon-charge-conversions and photon-spin-conversions are commensal processes. And these can be understood via the band structures in Fig. 2, in which the texture of $S_z$ in the energy window are rather homogeneous. On the other hand, the largest $\sigma_{Sz}^{y;xx}$ is -21 (nm·μA/V$^2$ $\hbar/2e$), corresponding to the shift-like photo-spin-conductance, and it is switched along with the reversion of magnetization.

Figures 3(e) and (f) show the photo-orbital-conductance. The highest value of injection-like $\sigma_{Lz}^{x;xx}$ is 6444 (nm·μA/V$^2$ $\hbar/e$) at 1.2 eV. $\sigma_{Lz}^{x;xx}$ is not reversed during the reversion of magnetization, similar to the case of photo-orbital-conductance. On the other hand, the shift-like photo-orbital-conductance $\sigma_{Lz}^{y;xx}$ has largest value -183 (nm·μA/V$^2$ $\hbar/e$) at 1.05 eV, and the sign of $\sigma_{Lz}^{y;xx}$ is changed when we reverse the magnetization. Due to the nontrivial orbital angular momentums texture in the energy window as displayed in Fig. 2, the evolution of photo-orbital-conductance is significantly different from the photoconductance and photo-spin-conductance. All these results indicate that the PT-broken is essential to manipulate the spin and orbital degrees of freedoms (DOF) in BPVEs.

To further understood the bulk photocurrents in momentum space, Fig. 4 displays the BZ distributions of photocurrents under $x$-polarized incident light with energy 1.05 eV. Figure 4(a) shows the injection-type photocurrents along $x$-direction. The distribution is highly asymmetric about the $y$-axis. The injection currents in left part of BZ with generally negative group velocities is much stronger than the ones in right part, leading to the intense photocurrents propagating along $-x$-direction. Since the breaking of mirror symmetry about the $y$-axis in 2H-FeCl$_2$ is totally due to the pseudo-vector nature of magnetic moments on Fe$^{2+}$ ions, thus the generation of large photocurrents should be mainly attributed to the magnetizations. Figure 4(b) also shows the photocurrents along $x$-direction, but the magnetization of 2H-FeCl$_2$ is reversed. The distribution of photocurrents is anti-symmetric to the case presented in Fig. 4(a). The states in right part of BZ with positive group velocities dominate the BPVE, and the net photocurrents propagate in +x-direction, in line with the results in Figs. 3(a) and (b).

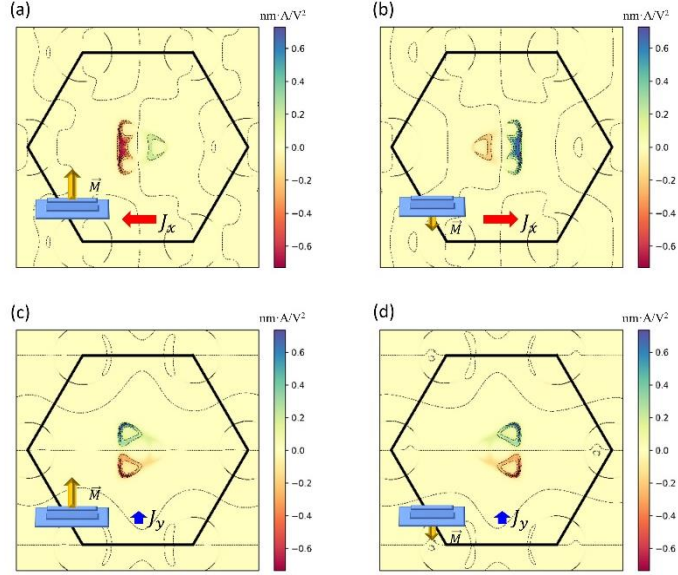

**Fig. 4** The distribution of photocurrents on BZ of 2H-FeCl$_2$ under illumination of x-polarized light with energy 1.05 eV. **(a, c)** The photocurrents along x-direction. **(b, d)** The photocurrents along y-direction. Insets depict the magnetization orientation of FeCl$_2$ in each case.

Figure 4(c) gives the shift-like photocurrents along y-direction. The states in up and down part of BZ contribute to the positive and negative photocurrents, respectively. The contribution of single Bloch state with certain crystal moment is comparable to the cases presented in Figs. 4(a) and (b), with maximal value ~0.6 (nm·A/V$^2$). However, the distributions on upper and lower parts of BZ are almost anti-symmetric, and the photocurrents are thus nearly compensated. For the case with reversed magnetization [Fig. 4(d)], the distribution of photocurrents is varied, but the whole shape is the same as before, justifying that the shift currents in y-direction are irrelevant to the magnetization orientation.

5. Discussion and conclusion

The 2H-FeCl$_2$ and BPVEs discussed here are different from the 1T-FeCl$_2$ reported previously [37, 60]. First, the 1T-FeCl$_2$ monolayer has inversion symmetry, [37] thus displays no BPVEs, but the 2H-FeCl$_2$ monolayer can host finite BPVEs. Second, the bilayer anti-ferromagnetic 1T-FeCl$_2$ possesses PT-symmetry and the BPVEs in bilayer 1T-FeCl$_2$ belong to the class 2, but the BPVEs in monolayer 2H-FeCl$_2$ belong to the class 3. Third, the working temperature (also the Nèel temperature) of BPVEs in bilayer 1T-FeCl$_2$ is ~25 K due to the weak interlayer exchange couplings, [60] but the BPVEs in monolayer 2H-FeCl$_2$ are supposed to work beyond room temperature [51] due to the high Curie temperature supported by the strong intralayer exchange couplings.

Therefore, the 2H-FeCl$_2$ is supposed to have better performance than the 1T-phase.

The thickness is usually relevant to the properties of thin film. Fortunately, due to the weak vdW interlayer interactions and the peculiar symmetry structure of 2H-FeCl$_2$, the thickness effects are minimal, and most of the properties of BPVEs in monolayers hold in the multi-layer case and even in the 3D case which involves periodicity in the thickness direction (See the supplementary information for more details).

All the three types of photo-induced currents can be readily detected via usual experiments methods as shown in Fig. 5. Since the photoconductance $\sigma_C^{x;xx}$ of 2H-FeCl$_2$ is large, the linearly polarized light induced photocurrents in circuits can be detected between the electrodes on zigzag edges [Fig. 5(a)]. The photocurrents measured between electrodes on armchair edges are much smaller due to the anisotropic nature of $\sigma_C$. These BPVE signals are shown under zero-bias-voltage condition, distinguishing from the conventional photocurrent responses of heterostructures. Besides, the device in Fig. 5(a) displays the prototype of 2D nonlinear solar cell, which is promising for the future energy applications. Furthermore, the current signals detected via device in Fig. 5(a) will be switched once the perpendicular magnetization reversed. So, this device can be used to read out the magnetic memory in 2H-FeCl$_2$.

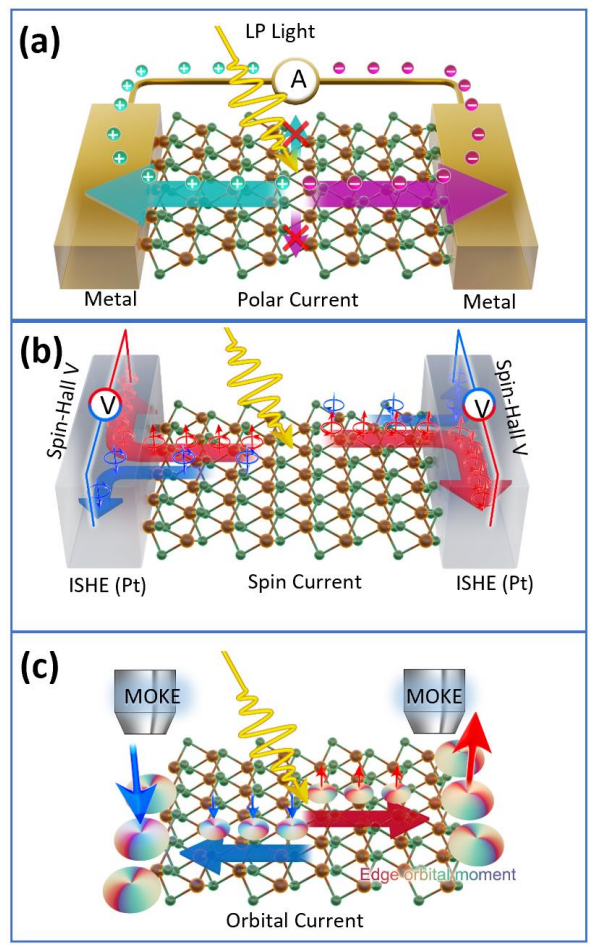

**Fig. 5** Detections of photocurrents, photo-spin-currents, and photo-orbital-currents in 2H-FeCl$_2$. **(a)** Detecting the photocurrents along *x*-axis via measuring the currents under zero-bias. The Green and purple arrows along

*x*-axis denote the flowing of electrons and holes induced by the illuminations of linearly polarized lights. The arrows along *y*-axis denote the unfavored propagation directions of photo-carriers. **(b)** Detecting the photo-spin-currents via measuring the Hall voltage produced by the ISHE. Red and blue arrows denote the flowing of spin-currents with spin-polarizations along +*z* and -*z* directions. **(c)** Detecting the orbital magnetic moments on edges via MOKE. The colored tori represent the orbital angular moments and the colors are the phases. The in-plane blue and red arrows denote the flowing of orbital angular moments. The out-of-plane arrows on edges denote the orbital magnetic moments caused by the accumulations of orbital moments on edges.

The inversed spin Hall effects (ISHE) can be utilized to measure the photo-spin-currents [Fig. 5(b)]. Once the photo-spin-currents inject into the electrodes with ISHE, e.g., Pt-electrodes, the spin-currents with opposite spin-polarizations are deflected to different transverse directions. The Hall voltage thus can be detected in the ISHE electrodes. Due to the highly anisotropic transporting of photo-spin-currents as we shown before, the electrodes placed on zigzag edges show significantly larger Hall voltages than those on armchair edges. Figure 5(c) shows the device utilized to detect the photo-orbital-currents. The orbital-currents propagate along the *x*-axis and the opposite orbital angular moments are accumulated in opposite zigzag edges, leading to the nonequilibrium edge orbital angular moments and orbital magnetic moments. Then one can directly measure these responses of magnetic moments via magneto-optical Kerr microscope (MOKE). The devices displayed in Figs. 5(b) and (c) are also the prototypes of 2D spintronics and orbitronics, offering opportunities to read/write nonvolatile information from/into the magnetic memories in the manner of low-consumptions.

In summary, we have shown the class 3 BPVEs in 2H-FeCl$_2$ using relativistic DFT and perturbation theory. The light-to-current, light-to-spin-current, and light-to-orbital-current conversions in monolayer 2H-FeCl$_2$ are robust due to the peculiar PT-broken symmetry and the strong intra-layer ferromagnetic interactions, which are stable up to room temperature. The injection-like currents propagate in *x*-axis, while the shift-like currents are in *y*-axis. The BPVE efficiency in 2H-FeCl$_2$ is large with photoconductance exceeding 4650 (nm·mA/V$^2$). More importantly, all the three types of currents can be flexibly switched via rotating the magnetization orientation of 2H-FeCl$_2$. The highly controllable and sizable 2D BPVEs in 2H-FeCl$_2$ make it an ideal platform to realize 2D energy, spintronic, and orbitronic applications.

**Declarations** The authors declare that they have no competing interests and there are no conflicts.

**Acknowledgements** This work was supported by the National Natural Science Foundation of China (Nos. 52275565 and 62075139), the Natural Science Foundation of Shandong Province (No. ZR2022QA019), the Natural Science Foundation of Guangdong (No. 2022A1515011667), the Youth Talent Fund of Guangdong Province (No. 2023A1515030292), Shenzhen Foundation Research Key Project (No. JCYJ20200109114244249), and Shenzhen Science and Technology Innovation Commission (No. RCJC20200714114435063).